%% file: main.tex
\documentclass[conference]{IEEEtran}
\IEEEoverridecommandlockouts

\usepackage{cite}
\usepackage{amsmath,amssymb,amsfonts}
\usepackage{algorithmic}
\usepackage{graphicx}
\usepackage{textcomp}
\usepackage{xcolor}
\usepackage{gensymb}
\usepackage{yhmath}
\usepackage{mathdots}
\usepackage{algorithm}
\usepackage{algorithmic}
\usepackage{bm}
\usepackage{soul}
\usepackage[caption=false,font=footnotesize]{subfig}

\newcommand{\nris}{N}
\newcommand{\nx}{N_{x}}
\newcommand{\ny}{N_{y}}
\newcommand{\nxt}{\tilde{N}_{x}}
\newcommand{\nyt}{\tilde{N}_{y}}
\newcommand{\snm}{\mathbf{s}_{n,m}}
\newcommand{\Kt}{\tilde{K}}

\newcommand{\rnmk}[1]{r_{n,m}^{#1}}
\newcommand{\UEk}[1]{\mathbf{u}_{#1}}
\newcommand{\ori}{\mathbf{g}(\psi,\gamma)}
\newcommand{\etp}{\mathbf{e}(\theta,\phi)}
\newcommand{\evec}{\mb{e}}
\newcommand{\bvec}{\mb{g}}
\newcommand{\dx}{d_x}
\newcommand{\dy}{d_y}

\newcommand{\delex}{\delta_{ex}}
\newcommand{\deley}{\delta_{ey}}
\newcommand{\delrk}[1]{\delta_{r,#1}}
\newcommand{\aik}[2]{\left[\mb{A}\right]_{#1,#2}}
\newcommand{\bik}[2]{\left[\mb{B}\right]_{#1,#2}}
\newcommand{\phase}[1]{\frac{j#1\pi}{\lambda}} 
\newcommand{\pinv}[1]{#1^{\dag}} 
\newcommand{\mb}[1]{\mathbf{#1}} 
\newcommand\norm[1]{\lVert#1\rVert} 

\def\BibTeX{{\rm B\kern-.05em{\sc i\kern-.025em b}\kern-.08em
    T\kern-.1667em\lower.7ex\hbox{E}\kern-.125emX}}
\begin{document}

\title{Near-field 5D Pose Estimation using Reconfigurable Intelligent Surfaces
}

\author{Srikar Sharma Sadhu,  Praful D. Mankar, and Santosh Nannuru
\thanks{S. S. Sadhu,  P. D. Mankar and S. Nannuru are with Signal Processing and Communication Research Center, IIIT Hyderabad, India. Email: srikar.sadhu@research.iiit.ac.in, \{praful.mankar,santosh.nannuru\}@iiit.ac.in.  
} 
}

\maketitle

\begin{abstract}
The advent of 6G is expected to enable many use cases which may rely on accurate knowledge of the location and orientation of user equipment (UE). The conventional localization methods suffer from limitations such as synchronization and high power consumption required for multiple active anchors. This can be mitigated by utilizing a large dimensional passive reconfigurable intelligent surface (RIS). This paper presents a novel low-complexity approach for the estimation of 5D pose (i.e. 3D location and 2D orientation) of a UE in near-field RIS-assisted multiple-input multiple-output (MIMO) systems. The proposed approach exploits the symmetric arrangement of uniform planar array of RIS and uniform linear array of UE to decouple the 5D problem into five 1D sub-problems. Further, we solve these sub-problems using a total least squares ESPRIT inspired approach to obtain closed-form solutions. 
\end{abstract}

\begin{IEEEkeywords}
Near-field localization, ESPRIT, RIS, 5D pose
\end{IEEEkeywords}\vspace{-2mm}
\section{Introduction} \label{sec:Introduction}\vspace{-1mm}
6G networks are envisioned to enable on-demand high-speed, low power and ultra low-latency communication  via leveraging the advancements in antenna array technologies involving  reconfigurable intelligent surfaces (RISs),  massive multiple-input multiple-output (MIMO), cell-free MIMO, etc. 
With such advancements, 6G is expected to facilitate unprecedented range of new use cases such as extended reality, online gaming, autonomous driving,  remote surgery,  industrial automation, etc., \cite{Giordani_2020,bourdoux20206gwhitepaperlocalization}. Many of these use cases by design  rely on sensing of various aspects including propagation environment,  user equipment (UE) location,  orientation of UE, etc.   
It is important to note that such sensing based services can  easily be facilitated using a base station (BS) assisted with passive RISs. This can potentially resolve important issues, like synchronization among multiple active reference nodes, that are usually required  for cooperation-based sensing methods. Inspired by this, our work focuses on the joint estimation of location and orientation of UE using  RIS-assisted MIMO communication systems.

{\em Related Works:}
In literature, localization has been heavily investigated for a variety of settings, particularly involving multiple cooperative anchors (equipped with antenna arrays) for performing UE triangulation using the estimates of received signal strength (RSS), angle of arrival (AoA),  time of arrival (ToA), etc. The  readers interested in this direction may refer to a few excellent related works \cite{shen2010accuracy,Wymeersch_Fundamental_Limits,Nil_2017} and survey articles \cite{Zafari_2019,Cheng_2012_survey} for more details. 
However, such localization methods requiring multiple active anchors have restricted applications in practical scenario because of 1) the need of cooperation among the anchor nodes, 2) the absence of line-of-sight (LoS) links, and 3) high power consumption.   
These limitations fortunately can be tackled to a large extent by utilizing  RISs as reference anchor nodes. The RIS is a low cost passive array consisting of many reflective elements that can be deployed large in number to navigate the signal around obstacles to ensure indirect LoS connectivity  \cite{Basar}. Such passive RISs can be utilized to assist a single anchor node for UE localization without requiring any cooperation from other active anchors \cite{Dardari_NLoS_2022}, which essentially can reduce the signaling overhead and power consumption. Because of these benefits,  there is an opportunity to enhance the ability of localization using RIS as the reference nodes \cite{Wymeersch_2020_ris,He_2022}. 

The recent works utilizing RIS for localization have primarily focused on the near-field scenario. This is because of  the rapid increase in both the operational frequency  and    antenna array dimensions to meet the  requirements  of modern communication networks. This in turn significantly alters the   physical characteristics of propagation environment, effectively  constraining the communication range to the near-field region  \cite{Mingyao_2023,Haiyang_2023_6g}. Besides, the received signal power of RIS-assisted indirect link is expected to be higher when the UE is placed in the near-field of RIS as compared to that in far-field scenario  because of their underlying path loss models. 
On this direction, the authors of  \cite{Dardari_NLoS_2022,Mingan_PhaseDesign_2022,Cuneyd_2023,Yijin_2023,Rinchi_2022} consider the near-field localization using a single anchor that is assisted by a single RIS. In particular, the authors of \cite{Dardari_NLoS_2022} focus on designing signaling and positioning methods while utilizing the time-varying reflection coefficients of the RIS to localize the UE in the absence of LoS link, which makes the proposed algorithms useful in harsh propagation conditions. In \cite{Mingan_PhaseDesign_2022}, the authors first determine the   squared position error bound (SPEB) and utilize it to  design the RIS phase response in two methods 1) by minimizing the average localization accuracy that is average of SPEB over area of interest (AOI)  and 2) minimizing the maximum of SPEB over AOI. 
Further, a low complex approximate mismatched maximum likelihood (ML) estimator is developed in \cite{Cuneyd_2023} for RIS assisted near-field localization that decouples the 3D search (for range, azimuth and elevation) problem into three 1D searches.
In \cite{Yijin_2023}, a near-field channel estimation and localization algorithm is proposed based on the  second-order Fresnel approximation of the near-field channel
model. For the approximated channel model, array covariance matrix is derived  to facilitate  the decoupling of the UE distances and AoAs. Next, the authors employ sub-space method based  1D searches to estimate the distances and AoAs. 
Further, \cite{Rinchi_2022} proposed an  algorithm for RIS-assisted localization in near-field with multipath scenario. The proposed algorithm alternates over two steps, 1) extraction of the UE location using compressive sensing (CS) and 2) optimally configuring the RIS phase shifts based on the extracted parameters, until the convergence criteria is met.

While the aforementioned works focus on localizing a single-antenna UE, a few recent works \cite{Shengqiang_2024,Arash_2018,Mohammad_2023,Jianxiu_2022,Ahmed_2021}  explore joint estimation of position and orientation, {\em termed as pose estimation}, of multi-antenna equipped UE, which can  be useful for enabling new use cases. 
The authors of \cite{Shengqiang_2024} shows the connection of AoA and angle of departure (AoD) with the projection model from computer vision and employ  perspective projection method for 6D pose estimation using multiple single-antenna BS. A two stage   pose estimation algorithm that achieves Cramer-Rao bound for mmWave MIMO system is developed in \cite{Arash_2018}. The first stage of the proposed algorithm applies CS-based vector matching pursuit method for the coarse estimation to exploit the sparsity of mmWave channel in the AoA and AoD domain, and  the second stage applies the space-alternating generalized expectation maximization for the fine refinement. Next, the authors of \cite{Mohammad_2023} constructed an ML framework for 6D pose estimation in a mmWave orthogonal frequency division multiplexing (OFDM)-MIMO downlink system  and then, to avoid complex exhaustive search,  proposed a  geometric ad-hoc estimator for parameter initialization which reduces the high-dimensional ML problem to 1D search over finite intervals. Further, the authors of \cite{Jianxiu_2022}, leveraging the inherent sparsity of mmWave channel for MIMO-OFDM system, proposed an atomic norm minimization framework to jointly estimate ToA, AoA and AoD. These estimates are further utilized with non-linear least squares method to estimate the UE pose. Furthermore, \cite{Ahmed_2021} considers near-field pose estimation using RIS to obtain the performance limits of pose estimation problem and then focuses on design of RIS phase shifts that enable joint communication and pose estimation. However, there is limited research on using RIS for pose estimation. Furthermore, most of these existing approaches are algorithmic or search-based in nature, which often leads to increased complexity of the overall system. 
 
This paper considers 5D pose estimation of UE equipped with uniform linear array (ULA) in near-field using RIS-assisted MIMO  systems. 
To obtain a low-complex solution, we perform various transformations to decouple this 5D pose estimation problem into five sub-problems by leveraging the geometric structure of near-field channel model for considered system. Further, we solve these sub-problems using total least squares (TLS) ESPRIT  algorithm \cite{Richard_1989_esprit} and obtain  closed-form estimates for all the pose parameters.

{\em Notations:}
Vectors are denoted using bold lowercase letters (e.g. $\mb{q}$) and matrices are denoted using bold uppercase letters (e.g. $\mb{Q}$).
Transpose, conjugate, conjugate transpose and pseudo-inverse are represented as $(.)^T$, $(.)^*$, $(.)^H$ and $\pinv{(.)}$, respectively.
The operator $\operatorname{diag}(\mb{q})$ represents a diagonal matrix with diagonal $\mb{q}$. 
The Khatri-Rao, Hadamard and Kronecker products are denoted using $\circ$, $\odot$ and $\otimes$ respectively. $\mb{I}_Z$ represents a $Z \times Z$ identity matrix and  $\mb{F}_Z$ represents a flip matrix with $(i,Z-i+1)$-th entry being 1 for $i=1,\dots,Z$ and remaining being $0$. The notations $[\mb{Q}]_{l,:}$ and $[\mb{Q}]_{:,l}$ respectively  denote $l$-th row and $l$-th column of matrix $\mb{Q}$. \vspace{-1mm}
\section{System Model}\label{sec:system}
\begin{figure}[t]
\centerline{\includegraphics[width=0.37\textwidth]{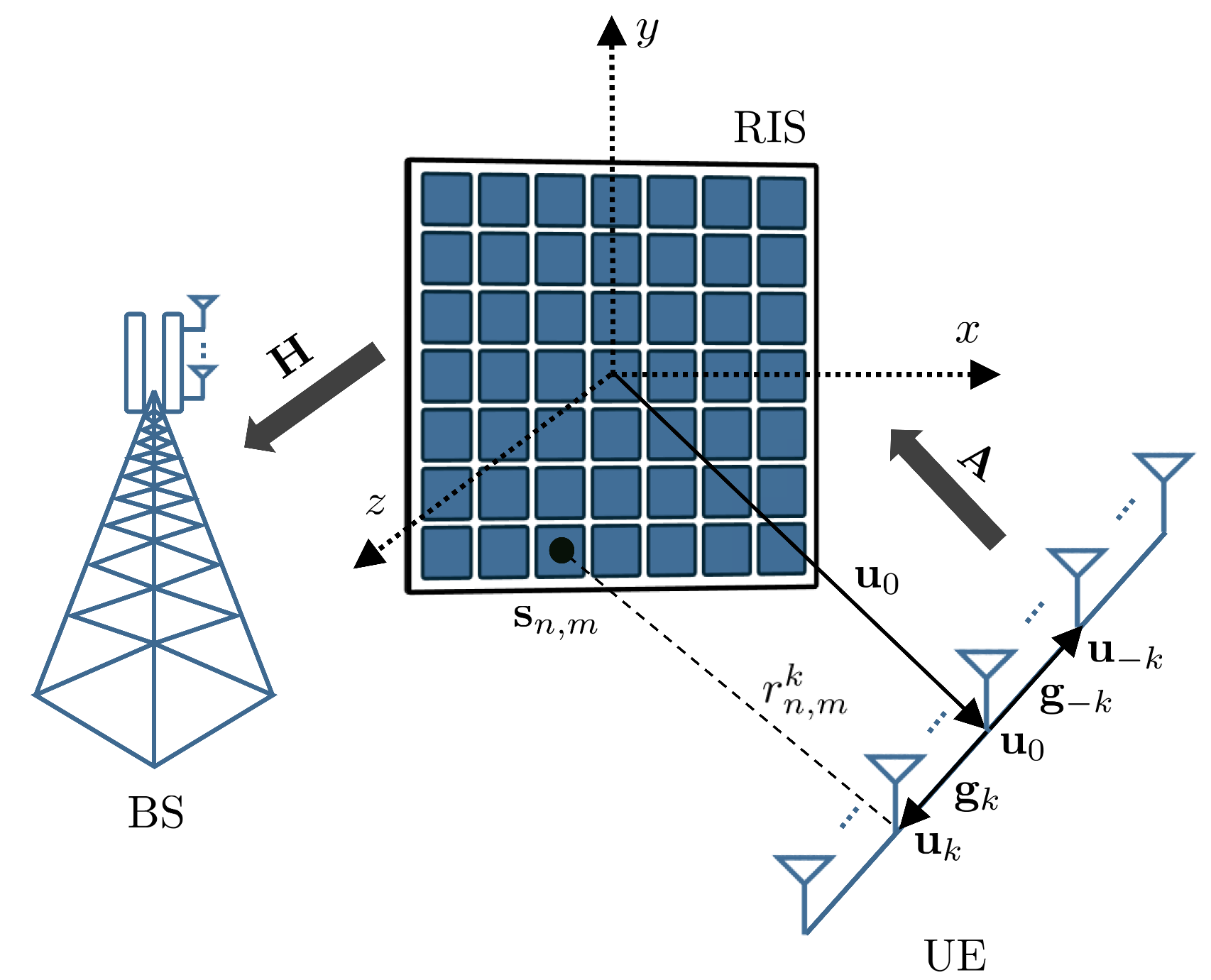}}\vspace{-4mm}
\caption{An Illustration of Considered System Model}
\label{fig:system_model}\vspace{-6mm}
\end{figure}

We consider an RIS-assisted MIMO wireless system, as shown in Fig \ref{fig:system_model}. The RIS is a uniform planar array (UPA) with $\nris$ elements. Without loss of generality, we assumed that the RIS is placed in the $xy$ plane centered at the origin such that $\nris  =  \nx\ny$ where $\nx=2\nxt+1$ and $\ny=2\nyt+1$ represent the number of elements along the $x$ and $y$ axes, respectively. The inter-element spacings along the $x$ and $y$ axes are considered to be $\dx$ and $\dy$. 
Thus, the location of $(n,m)$-th element of RIS becomes $\snm = \left[n\dx,m\dy,0\right]^T$ for $n \in \mathbb{N} = \{-\nxt,\ldots,0,\ldots,\nxt\}$ and $m \in \mathbb{M} = \{-\nyt,\ldots,0,\ldots,\nyt\}$.
\vspace{-3.5mm}

The UE location $\UEk{0}=r\etp$ is considered to lie within the near-field of the RIS, where $r$, $\theta$, and $\phi$ denote the distance, azimuth angle and elevation angle  with respect to the origin (i.e., center of RIS), respectively, and $\etp$ is  given by 
\begin{align}
        \etp &\triangleq \left[\cos\theta\cos\phi,~\sin\theta\cos\phi,~\sin\phi\right]^T.\label{eq:e_def}
\end{align}
The UE is equipped with a ULA comprising of $K = 2\Kt + 1$ antennas which are placed symmetrically across the UE location $\UEk{0}$ with orientation $\ori$ given as 
\begin{align}
    \ori &\triangleq \left[\cos\psi\cos\gamma,~\sin\psi\cos\gamma,~\sin\gamma\right]^T,\nonumber
\end{align}
where $\psi$ and $\gamma$ are the azimuth angle and elevation angle of orientation with respect to $\UEk{0}$. 
Therefore, the $k$-th antenna element of the ULA can be determined as
\begin{equation}
    \UEk{k} = \UEk{0} + kd_u\ori,\nonumber
\end{equation}
$\forall k \in \mathbb{K} = \{-\Kt,\ldots,0,\ldots,\Kt\}$ 
where $d_u$ is the inter-antenna spacing. 
For ease of notation, $\etp$ and $\ori$ will be shortened to $\evec$ and $\bvec$, respectively.

We consider the RIS-assisted uplink scenario wherein the UE transmits a reference signal  and the  BS receives it through the RIS to localize the UE. 
The direct link between BS and UE is assumed to be absent. The indirect link between BS and UE through RIS involves UE-RIS and RIS-BS links, both of which are assumed to LoS links.
The reference transmit sequence  $\mb{S}\in\mathbb{C}^{K\times L}$ 
is constructed such that $\mb{S}\mb{S}^H=P_TK^{-1}\mb{I}_K$, where $L$ represents the number of transmissions. 
The vector response of the near-field channel between $k$-th antenna at the UE and the RIS is
\begin{equation}
    \mb{a}_{k} = \left[e^{-j2\pi \left(r_{-\nxt,-\nyt}^{k} - ~r\right)/\lambda},\dots,e^{-j2\pi \left(r_{\nxt,\nyt}^{k} - ~r\right)/\lambda}\right]^T,\label{eq:a_k}\nonumber
\end{equation}
for $k \in \mathbb{K}$, where $\rnmk{k} = \norm{\UEk{k} - \snm}$ is the distance between the $k$-th antenna at UE and $(n,m)$-th element of RIS. 
The LoS link from UE to RIS is modeled using an $N\times K$ matrix as
\begin{equation}\label{eq:A}
    \mb{A} = \left[\mb{a}_{-\Kt}, \dots, \mb{a}_{0}, \dots, \mb{a}_{\Kt}\right].
\end{equation}
We consider the BS to be located in the far-field of the RIS and model the  RIS-BS link  using an $M\times N$ matrix as
\vspace{-1mm}
\begin{equation}\label{eq:H}
    \mb{H} = \mb{h}_{b} \left(\mb{h}_{rx} \otimes \mb{h}_{ry}\right)^H,
\end{equation}
where $\mb{h}_{b} = \mb{h}(M,d_{b}\sin\theta_B)$, $\mb{h}_{rx}= \mb{h}(N_x,d_{x}\cos\theta_R\cos\phi_R)$ and $\mb{h}_{ry}= \mb{h}(N_y,d_{y}\sin\theta_R\cos\phi_R)$
such that 
$\mb{h}(T,\zeta) = \left[e^{j2\pi\tilde{T} \zeta/\lambda},\dots,1,\dots, e^{-j2\pi \tilde{T} \zeta/\lambda}\right]^T$,
and $\tilde{T}=\frac{T-1}{2}$.
Here, $\theta_B$ is the angle of arrival at the BS, $d_{b}$ is the inter-antenna spacing and $\theta_R$ and $\phi_R$ are the azimuth and elevation angles of departure from the RIS. The signal is received at the BS under $P$ RIS phase shift configurations. The RIS configuration matrix $\mb{\Phi} \in \mathbb{C}^{P \times N}$ is designed with $(p,i)$-th element as $[\mb{\Phi}]_{p,i} = \exp\left(-j2\pi\left(p-1\right)\left(\frac{i-1}{\nris}\right)\right)$ to ensure that all $P$ configurations are orthogonal. The RIS phase shift matrix under $p$-th configuration is given by $\mb{\Omega}_p = \operatorname{diag} \left([\mb{\Phi}]_{p,:}\right)$.
From \eqref{eq:A} and \eqref{eq:H}, the signal received at the BS for $l$-th transmission under $p$-th RIS configuration can be written as
\begin{equation}\label{eq:rec_signal}
    \mb{y}_{p}(l) = \mb{H}\mb{\Omega}_p\mb{A}[\mb{S}]_{:,l} + \mb{w}_{p}(l),
\end{equation}
where $\mb{w}_p(l)$ is zero mean complex Gaussian noise with covariance matrix $\sigma^2\mb{I}_M$. For $p$-th configuration, we define an observation matrix $\mb{Y}_p \triangleq \left[ \mb{y}_p(1) \ldots \mb{y}_p(L) \right]$ and the noise matrix $\mb{W}_p \triangleq \left[ \mb{w}_p(1) \ldots \mb{w}_p(L) \right]$. Using \eqref{eq:rec_signal}, we can write
\begin{equation}\label{eq:Yp}
    \mb{Y}_p = \mb{H}\mb{\Omega}_p\mb{A}\mb{S} + \mb{W}_{p}.
\end{equation}
By column-wise stacking of ${M\times L}$ observation matrices $\mb{Y}_p$ for $p=1,\dots,P$, we construct an ${MP\times L}$ matrix as
\begin{equation}
    \mb{Y}=\bar{\mb{H}}\mb{A}\mb{S} + \bar{\mb{W}},\label{eq:obs_mat_stack}
\end{equation}
where $\bar{\mb{H}} = \left( \mb{\Phi} \circ \mb{H} \right)$ and $\bar{\mb{W}}$ is obtained by column-wise stacking of $\mb{W}_{p}$.
Note that the channel matrix $\mb{A}$ of UE-RIS link depends on the 5D pose parameters. 
Using \eqref{eq:obs_mat_stack}, a noisy version of $\mathbf{A}$ can be obtained as
\begin{align}\label{eq:A_ddot}
    \ddot{\mb{A}} = \pinv{\bar{\mb{H}}}\mb{Y}\pinv{\mb{S}}  = \mb{A} + \tilde{\mb{W}},
\end{align}
where $\tilde{\mb{W}} = \pinv{\bar{\mb{H}}}\bar{\mb{W}}\pinv{\mb{S}}$. To obtain \eqref{eq:A_ddot}, we need $P \geq N$ for ensuring  $\pinv{\bar{\mb{H}}}\bar{\mb{H}} = \mb{I}_{N}$. 

\section{5D User Pose Estimation}\label{sec:user-pose-estimation}
The aim is to estimate the 5D user pose $(\UEk{0},\bvec)$ which is a function of $(r,\theta,\phi,\psi,\gamma)$. 
The estimation of these parameters is coupled with each other. However, by exploiting the symmetrical arrangements of antenna arrays of RIS and UE, we could reformulate this multi-parameter estimation problem as multiple sub-problems, which are presented in the following subsections. These sub-problems are solved using TLS-ESPRIT inspired approach  \cite{Richard_1989_esprit}, leading to closed form expressions for the estimates of all five parameters.

\subsubsection{Estimation of Distance}\label{sec:distance_estimation}
Here, we present an approach to estimate the distance parameter $r$ using \eqref{eq:A_ddot} without depending on the angular parameters $(\theta,\phi,\psi,\gamma)$.
The distance between the UE's $k$-th  antenna and the RIS's $(n,m)$-th element is 
\begin{align}
    \rnmk{k} &= \norm{\UEk{k} - \snm} = \norm{r\evec + kd_u\bvec - \snm}.\nonumber
\end{align}
Using Fresnel approximation for the near-field \cite{Yijin_2023}, we approximate $\rnmk{k}$ as
{\small \begin{align}
    \rnmk{k} = r &+ \frac{\left(kd_u\right)^2 + \norm{\snm}^2}{2r} + kd_u(\evec^T\bvec - \frac{\bvec^T\snm}{r}) -\evec^T\snm.\nonumber
\end{align}}
Using this, the $(i,k)$-th element of  $\mb{A}$ can be rewritten as
\begin{equation}
    [\mb{A}]_{i,k}= e^{-\phase{2}\left(\frac{\left(kd_u\right)^2 + \norm{\snm}^2}{2r} + kd_u\evec^T\bvec - \frac{kd_u\bvec^T\snm}{r} -\evec^T\snm\right)},\nonumber
\end{equation}
where $i\triangleq(n+\nxt)\ny+(m+\nyt)+1$ for $n \in \mathbb{N}$ and $m \in \mathbb{M}$. 
Let $\mb{B}$ represents a matrix whose $(i,k)$-th element is defined as $[\mb{B}]_{i,k}\triangleq \aik{i}{k} \aik{i}{-k}$. From this, we get 
\begin{align}\label{eq:B_ik}
    \bik{i}{k} &= e^{\left(-\phase{4}\left(\frac{\left(kd_u\right)^2 + \norm{\snm}^2}{2r} - \evec^T\snm\right)\right)}.
\end{align}
Observe that $\bik{i}{k}$ is independent of orientation parameters $\{\psi,\gamma\}$. 
We can represent $\mb{B}$ in matrix form as
\begin{equation}
    \mb{B} = \mb{A} \odot \left(\mb{A}\mb{F}_{K}\right),  \nonumber  
\end{equation}
where $\mb{F}_K$ is the flip matrix as defined under notations. 

Now, we exploit the  geometries of RIS and UE antenna arrays  to identify the phase difference in the signal observed at successive elements in terms of parameter $r$ only. Further, similar to  ESPRIT algorithm \cite{Richard_1989_esprit}, we use the knowledge of such phase difference to estimate the parameter $r$.
Note that the phase difference between the elements $\bik{i}{k}$ and $\bik{i}{k+1}$ can be obtained using \eqref{eq:B_ik} only in terms of parameter $r$ as
\begin{equation}\label{eq:shift_r}
    \bik{i}{k+1}=\bik{i}{k}\delrk{k}
\end{equation}
where $\delrk{k}$ is the $k$-th entry of  $\bm{\delta}_{r} \in \mathbb{C}^{(K - 1)}$ and is given by
\begin{equation}\label{eq:delta_r_k}
    \delrk{k} = \exp\left(-\phase{2}\left(\frac{\left(2k + 1\right)d_{u}^{2}}{r}\right)\right).
\end{equation}
Then, using \eqref{eq:shift_r} and \eqref{eq:delta_r_k}, we can write
\begin{align}
    \mb{B}\mb{J}_{2r} &= \mb{B}\mb{J}_{1r}\operatorname{diag}\left(\bm{\delta}_{r}\right), \label{eq:esprit_r}
\end{align}
where $\mb{J}_{2r} = [\mb{0}_{K-1}|\mb{I}_{K-1}]^T$ and $\mb{J}_{1r} = [\mb{I}_{K-1}|\mb{0}_{K-1}]^T$ are the column selection matrices such that $\mb{0}_{K-1}$ is a zero-vector of length $K-1$. We rewrite \eqref{eq:esprit_r} for each
$k \in \mathbb{K} \setminus \{\Kt\}$ as
\begin{equation}
    \mb{b}_{k+1} = \mb{b}_{k}\delrk{k}.\nonumber
\end{equation}
Applying similar flip operation to \eqref{eq:A_ddot}, we get
\begin{equation}\label{eq:B_ddot}
     \ddot{\mb{B}}= \ddot{\mb{A}} \odot \left(\ddot{\mb{A}} \mb{F}_K\right) = \mb{B} + \tilde{\mb{B}},
\end{equation}
where {\small $\tilde{\mb{B}} =  \mb{A} \odot \left(\tilde{\mb{W}}\mb{F}_{K}\right) +\tilde{\mb{W}} \odot \left(\mb{A}\mb{F}_{K}\right) + \tilde{\mb{W}} \odot \left(\tilde{\mb{W}}\mb{F}_{K}\right)$}. From  \eqref{eq:esprit_r}, the estimation of $\bm{\delta}_{r}$ can be written in TLS form  \cite{Gene_2013_matrix} as   
\begin{equation}
    \ddot{\mb{B}}\mb{J}_{2r} = \ddot{\mb{B}}\mb{J}_{1r}\operatorname{diag}\left(\bm{\delta}_{r}\right). \notag
\end{equation}
Let $\ddot{\mb{B}} = [\ddot{\mb{b}}_{-\Kt},\dots,\ddot{\mb{b}}_{\Kt}]$. Using \eqref{eq:B_ddot}, we can relate  $\ddot{\mb{b}}_{k}$ and $\ddot{\mb{b}}_{k+1}$, i.e. noisy versions of $\mb{b}_{k}$ and $\mb{b}_{k+1}$, as
 \begin{equation}\label{eq:tls_delta_rk}
     \ddot{\mb{b}}_{k+1}= \ddot{\mb{b}}_k \delta_{r,k} \Rightarrow \mb{b}_{k+1} + \tilde{\mb{b}}_{k+1} = \left(\mb{b}_{k}+ \tilde{\mb{b}}_{k}\right)\delrk{k}.
 \end{equation}
Using  \cite[Sec. V-E]{Richard_1989_esprit}, we get a closed-form estimate of  {\small$\delrk{k}$}   as
\begin{equation}\label{eq:delta_rk_hat}
    \hat{\delta}_{r,k} = -\left[\mb{V}_{r,k}\right]_{1,2}\left(\left[\mb{V}_{r,k}\right]_{2,2}\right)^{-1},
\end{equation}
where $\mb{V}_{r,k} \in \mathbb{C}^{2 \times 2}$ are the left singular vectors of $\bm{\Delta}_{r,k} = \left[ \ddot{\mb{b}}_{k} | \ddot{\mb{b}}_{k+1}\right] \in \mathbb{C}^{\nris \times 2}$.
Finally, using the estimates $\hat{\delta}_{r,k}$ and \eqref{eq:delta_r_k}, we  obtain the estimate of parameter $r$ as
\begin{equation}\label{eq:r_hat}
    \hat{r} = \frac{1}{K-1}\sum\nolimits_{k = -\Kt}^{\Kt - 1} \frac{-2\pi \left(2{k}+1\right)d_{u}^{2}}{\lambda \angle \hat{\delta}_{r,k}}.
\end{equation}
\subsubsection{Estimation of Direction}\label{sec:direction_estimation}
In this subsection, we provide an approach to estimate  parameters of UE direction $(\theta,\phi)$  without any dependency on other parameters $(r,\psi,\gamma)$. For this, we first perform a few transformations on \eqref{eq:A_ddot} to remove the dependency and then apply a similar TLS-ESPRIT approach as presented in Section \ref{sec:distance_estimation}.

Let us define matrix $\mb{C}$ as 
\vspace{-1mm}
\begin{equation}
    \mb{C}=\mb{A}\odot\left(\mb{F}_N\mb{A}^*\mb{F}_K\right).\nonumber
\end{equation}
\vspace{-1mm}
The $(i,k)$-th element of $\mb{C}$ is given by
\begin{align}
    [\mb{C}]_{i,k}&=[\mb{A}]_{i,k}[\mb{A}^*]_{i_f,-k}=e^{\left( \phase{4}\left[\evec^T\snm-kd_u\mb{e}^T\mb{b}\right]\right)},\nonumber
\end{align}
such that  $i_f =(-n+\nxt)(\ny)+(-m+\nyt)+1$. Substituting vector $\evec$ given in \eqref{eq:e_def} and $\mb{s}_{n,m}$, we get 
\begin{align*}
     [\mb{C}]_{i,k}&=\left(\delex\right)^n\left(\deley\right)^m e^{\left( -\phase{4}kd_u\mb{e}^T\mb{b}\right)},\nonumber\\
    \text{such that}~~  \delta_{ex}&=\exp\left(\phase{4}d_x\cos{\theta}\cos{\phi}\right),\\
    \text{and}~~\delta_{ey}&=\exp\left(\phase{4}d_y\sin{\theta}\cos{\phi}\right).
\end{align*}
We can determine the phase difference of $[\mb{C}]_{i,k}$ with its adjacent horizontal and vertical entries corresponding to  RIS elements as
 \begin{align}\label{eq:delta_exy}
    [\mb{C}]_{i,k} &=  [\mb{C}]_{i_x,k}(\delta_{ex})^*=[\mb{C}]_{i_y,k}(\delta_{ey})^*,
\end{align}
 where $i_x=i+N_y$, $i_y=i+1$. 
 Now, we establish relations similar to \eqref{eq:esprit_r} for facilitating the estimation of parameters $(\theta,\phi)$ using TLS-ESPRIT approach. 
For this, we define the row selection matrices as
\begin{align}
     \mb{J}_{x1} &= [\mb{I}_{\nx(\ny - 1)}|\mb{0}_{\nris}]^T
     \text{~and~}\mathbf{J}_{y1}=\mb{I}_{N_x}\otimes[\mb{I}_{N_y-1}|\mb{0}_{N_y-1}],\nonumber\\
    \mb{J}_{x2} &= [\mb{0}_{\nris}|\mb{I}_{\nx(\ny - 1)}]^T\text{~and~}
     \mathbf{J}_{y2}=\mb{I}_{N_x}\otimes[\mb{0}_{N_y-1}|\mb{I}_{N_y-1}].\nonumber
\end{align}
Using these matrices, the relations in \eqref{eq:delta_exy} can be written  as
\begin{align*}
    \mb{J}_{x2}\mb{C} &= \mb{J}_{x1}\mb{C}\delta_{ex},\text{~and~} \mb{J}_{y2}\mb{C} = \mb{J}_{y1}\mb{C}\delta_{ey}.
\end{align*}
Applying a similar transformation to \eqref{eq:A_ddot}, we obtain
\begin{align}
    \ddot{\mb{C}} = \ddot{\mb{A}}\odot\left(\mb{F}_N\ddot{\mb{A}}^*\mb{F}_K\right)=\mb{C} +\tilde{\mb{C}},\label{eq:C_ddot}
\end{align}
where {\small $\tilde{\mb{C}}=\mb{A}\odot\left(\mb{F}_N\tilde{\mb{W}}^*\mb{F}_K\right)+\tilde{\mb{W}}\odot\left(\mb{F}_N\mb{A}^*\mb{F}_K\right)+\tilde{\mb{W}}\odot\left(\mb{F}_N\tilde{\mb{W}}^*\mb{F}_K\right)$}. Further, we obtain the following relations
\begin{align*}
    \mb{J}_{x2}\ddot{\mb{C}} &= \mb{J}_{x1}\ddot{\mb{C}}\delta_{ex}, \text{~and~}    \mb{J}_{y2}\ddot{\mb{C}} = \mb{J}_{y1}\ddot{\mb{C}}\delta_{ey}.
\end{align*}
Next, utilizing these relations, we estimate the phase differences $\delex$ and 
$\deley$, which can further be used to estimate the parameters $(\theta,\phi)$. 
Let $\ddot{\mb{C}}=[\ddot{\mb{c}}_{-\Kt},\dots,\ddot{\mb{c}}_{\Kt}]$.
Applying TLS, similar to \eqref{eq:tls_delta_rk}, to the following   column-wise relations
\begin{align*}
    \mb{J}_{x2}\ddot{\mb{c}}_{k}&= \mb{J}_{x1}\ddot{\mb{c}}_k \text{~and~}
    \mb{J}_{y2}\ddot{\mb{c}}_{k}= \mb{J}_{y1}\ddot{\mb{c}}_k \delta_{ex}
\end{align*}
provides the estimates of $\delex$ and $\deley$ in closed-forms as
\begin{align}
    \hat{\delta}_{ex} &= -\frac{1}{K}\sum\nolimits_{k\in\mathbb{K}}\left[\mb{V}_{ex,k}\right]_{1,2}\left(\left[\mb{V}_{ex,k}\right]_{2,2}\right)^{-1}, \label{eq:delta_ex_hat}\\
    \hat{\delta}_{ey} &= -\frac{1}{K}\sum\nolimits_{k\in\mathbb{K}}\left[\mb{V}_{ey,k}\right]_{1,2}\left(\left[\mb{V}_{ey,k}\right]_{2,2}\right)^{-1}, \label{eq:delta_ey_hat}
\end{align}
where  $\mb{V}_{ex,k}$ and $\mb{V}_{ey,k}$ are $2\times 2$  left singular vectors of $$\bm{\Delta}_{ex,k} = \left[\mb{J}_{x1}\ddot{\mb{c}}_{k} | \mb{J}_{x2}\ddot{\mb{c}}_{k} \right] \text{~and~} 
\bm{\Delta}_{ey,k} = \left[ \mb{J}_{y1}\ddot{\mb{c}}_{k} | \mb{J}_{y2}\ddot{\mb{c}}_{k}\right],$$
respectively.
 Further, using \eqref{eq:delta_ex_hat} and \eqref{eq:delta_ey_hat}, we can obtain
\begin{align}
    \cos{\hat\theta}\cos{\hat\phi}&={\lambda}({4\pi d_x})^{-1}\angle\hat{\delta}_{ex},\text{~and~}\label{eq:thetaphi_hat1}\\
    \sin{\hat\theta}\cos{\hat\phi}&={\lambda}({4\pi d_y})^{-1}\angle\hat{\delta}_{ey}. \label{eq:thetaphi_hat2}
\end{align}
This finally allows us to estimate the parameters $(\theta,\phi)$ as 
\begin{align}
    \hat{\theta} &\stackrel{(a)}{=} \arctan{\left(\frac{d_x\angle\hat{\delta}_{ey}}{d_y\angle\hat{\delta}_{ex}}\right)}, \label{eq:theta_hat}\\
    \hat{\phi} &\stackrel{(b)}{=} \arccos{\left(\frac{\lambda}{4\pi}\sqrt{\frac{(\angle\hat\delta_{ex})^2}{d_x^2} + {\frac{(\angle\hat{\delta}_{ey})^2}{d_y^2}}}\right)},\label{eq:phi_hat}
\end{align}
where step (a) is obtained by dividing \eqref{eq:thetaphi_hat2} with \eqref{eq:thetaphi_hat1},  whereas the step (b) is obtained by sum of squares of \eqref{eq:thetaphi_hat1} and \eqref{eq:thetaphi_hat2}.
\subsubsection{Estimation of Orientation} \label{sec:orientation_estimation}
Here, we present an approach for estimating the orientation parameters $(\psi,\gamma)$.  Unfortunately, it is difficult to decouple  these parameters from the location parameters $(r,\theta,\phi)$. Hence, we will use the estimates of these dependent parameters for facilitating the estimation of $(\psi,\gamma)$.

Let us define matrix $\mb{D}$ as
$$\mb{D}=\mb{A} \odot (\mb{F}_{\nris}\mb{A}^*),$$
such that its $(i,k)$-th element 
becomes  
\begin{align}
    [\mb{D}]_{i,k}= \aik{i}{k}[\mb{A}^{*}]_{i_f,k}  =e^{\left(\phase{4}\left(\frac{kd_u\bvec^T\snm}{r} + \evec^T\snm\right)\right)}.\nonumber    
\end{align}
where $i_f =(-n+\nxt)(\ny)+(-m+\nyt)+1$.
 The phase differences of $(i,k)$-th element of $\mb{D}$ with its adjacent horizontal and vertical entries corresponding to  RIS elements as
 \begin{align}\label{eq:delta_gxy}
    [\mb{D}]_{i,k} &=  [\mb{D}]_{i_x,k}(\delta_{gx,k})^*=[\mb{D}]_{i_y,k}(\delta_{gy,k})^*,
\end{align}
 where $i_x=i+N_y$, $i_y=i+1$ and 
 \begin{align}
    \delta_{gx,k} &= \delta_{ex} \exp{\left(\phase{4}\frac{kd_ud_x}{r}\cos{\psi}\cos{\gamma}\right)},\nonumber\\
    \delta_{gy,k} &= \delta_{ey} \exp{\left(\phase{4}\frac{kd_ud_y}{r}\sin{\psi}\cos{\gamma}\right)},\nonumber
\end{align}
such that $\delex$ and $\deley$ are given in Section \ref{sec:direction_estimation}.
Further, using the selection matrices constructed in Section \ref{sec:direction_estimation}, the relations in \eqref{eq:delta_gxy} can be written in matrix forms as
\begin{align*}
    \mb{J}_{x2}\mb{D} &= \mb{J}_{x1}\mb{D}\operatorname{diag}(\bm{\delta}_{gx}),\text{~and~}\mb{J}_{y2}\mb{D} = \mb{J}_{y1}\mb{D}\operatorname{diag}(\bm{\delta}_{gy}),
\end{align*}
where \mbox{\small $\bm{\delta}_{gx}=[\delta_{gx,-\Kt},\dots,\delta_{gx,\Kt}]$ and $\bm{\delta}_{gy}=[\delta_{gy,-\Kt},\dots,\delta_{gy,\Kt}]$}.

Applying similar  transformation to \eqref{eq:A_ddot} gives
\begin{align}\label{eq:D_ddot}
        \ddot{\mb{D}} = \ddot{\mb{A}} \odot \left(\mb{F}_{\nris}\ddot{\mb{A}}^*\right) = \mb{D} + \tilde{\mb{D}},
\end{align}
where 
{\small $\tilde{\mb{D}} = \mb{A} \odot \left(\mb{F}_{\nris}\tilde{\mb{W}}^*\right) + \tilde{\mb{W}} \odot \left(\mb{F}_{\nris}\mb{A}^*\right) + \tilde{\mb{W}} \odot \left(\mb{F}_{\nris}\tilde{\mb{W}}^*\right)$}.
Further, we obtain the following relations  
\begin{align*}
    \mb{J}_{x2}\ddot{\mb{D}} &= \mb{J}_{x1}\ddot{\mb{D}}\operatorname{diag}(\bm{\delta}_{gx}),\text{~and~}\mb{J}_{y2}\ddot{\mb{D}} = \mb{J}_{y1}\ddot{\mb{D}}\operatorname{diag}(\bm{\delta}_{gy}).
\end{align*}
Let $\ddot{\mb{D}}=[\ddot{\mb{d}}_{-\Kt},\dots,\ddot{\mb{d}}_{\Kt}]$.
Applying TLS-ESPRIT to the following column-wise relations
\begin{algorithm}[b]
    \caption{TLS-ESPRIT-based 5D User Pose Estimation}
    \label{alg:5d_pose_estimation}
    \begin{algorithmic}[1]
        \REQUIRE $\mb{Y}$, $\mb{S}$, $\mb{H}$, $\mb{\Phi}$, $\lambda$, $M$, $N$, $K$.
        \ENSURE Estimated user pose: $(\hat{r}, \hat{\theta}, \hat{\phi}, \hat{\psi}, \hat{\gamma})$

        \STATE Compute $\ddot{\mb{A}}$ using $\mb{Y}$ and \eqref{eq:A_ddot}
        
        \STATE Compute $\ddot{\mb{B}}$ using $\ddot{\mb{A}}$ and \eqref{eq:B_ddot}
        \STATE Obtain $\hat{r}$ using \eqref{eq:r_hat}
        
        \STATE Compute $\ddot{\mb{C}}$ using $\ddot{\mb{A}}$ and \eqref{eq:C_ddot}
        \STATE Obtain $\hat{\theta}$ using \eqref{eq:theta_hat}
        \STATE Obtain $\hat{\phi}$ using \eqref{eq:phi_hat}
        
        \STATE Compute $\ddot{\mb{D}}$ using $\ddot{\mb{A}}$ and \eqref{eq:D_ddot}
        \STATE Obtain $\hat{\psi}$ using \eqref{eq:psi_hat}
        \STATE Obtain $\hat{\gamma}$ using \eqref{eq:gamma_hat}
        
        \RETURN $(\hat{r}, \hat{\theta}, \hat{\phi}, \hat{\psi}, \hat{\gamma})$
    \end{algorithmic}
\end{algorithm}
\setlength{\textfloatsep}{8.5pt}
\begin{align*}
    \mb{J}_{x2}\ddot{\mb{d}}_{k}&= \mb{J}_{x1}\ddot{\mb{d}}_k \delta_{gx,k},\text{~and~}\mb{J}_{y2}\ddot{\mb{d}}_{k}= \mb{J}_{y1}\ddot{\mb{d}}_k \delta_{gy,k},
\end{align*}
provides the estimates of $\delta_{gy,k}$ and $\delta_{gy,k}$ in closed-forms as
\begin{align}
    \hat{\delta}_{gx,k} &= -\left[\mb{V}_{gx,k}\right]_{1,2}\left(\left[\mb{V}_{gx,k}\right]_{2,2}\right)^{-1},\text{~and}\label{eq:delta_gx_hat}\\
    \hat{\delta}_{gy,k} &= -\left[\mb{V}_{gy,k}\right]_{1,2}\left(\left[\mb{V}_{gy,k}\right]_{2,2}\right)^{-1},\label{eq:delta_gy_hat}
\end{align}
where  $\mb{V}_{gx,k}$ and $\mb{V}_{gy,k}$ are $2\times 2$  left singular vectors of $$\bm{\Delta}_{gx,k} = \left[\mb{J}_{x1}\ddot{\mb{d}}_{k} | \mb{J}_{x2}\ddot{\mb{d}}_{k} \right] \text{~and~} 
\bm{\Delta}_{gy,k} = \left[ \mb{J}_{y1}\ddot{\mb{d}}_{k} | \mb{J}_{y2}\ddot{\mb{d}}_{k}\right],$$
respectively. Using \eqref{eq:delta_gx_hat} and \eqref{eq:delta_gy_hat} along with \eqref{eq:delta_gxy}, we obtain
\begin{align}
    \cos{\hat\psi}\cos{\hat\gamma}&={\lambda \hat{r}}({4\pi kd_ud_x})^{-1}\alpha_{x,k},\text{~and~}\label{eq:psigamma_hat1}\\
    \sin{\hat\psi}\cos{\hat\gamma}&={\lambda \hat{r}}({4\pi kd_ud_y})^{-1}\alpha_{y,k}, \label{eq:psigamma_hat2}
\end{align}
where $\alpha_{x,k} = \angle(\hat{\delta}_{gx,k}\hat{\delta}_{ex}^{-1})$ and $\alpha_{y,k} = \angle(\hat{\delta}_{gy,k}\hat{\delta}_{ey}^{-1})$.
Finally, using \eqref{eq:psigamma_hat1} and \eqref{eq:psigamma_hat2}, we  estimate the orientation parameters  as 
\begin{align}
    \hat{\psi} &\stackrel{(c)}{=} \frac{1}{K - 1} \sum_{k \in \mathbb{K} \setminus \{0\}} \arctan{\left(\frac{d_x\alpha_{y,k}}{d_y\alpha_{x,k}}\right)}, \label{eq:psi_hat}\\
   \hat{\gamma} &\stackrel{(d)}{=} \frac{1}{K - 1} \sum_{k \in \mathbb{K} \setminus \{0\}} \arccos{\left(\frac{\lambda \hat{r}}{4\pi kd_u}\sqrt{\frac{\alpha_{x,k}^2}{d_x^2} + \frac{\alpha_{y,k}^2}{d_y^2}}\right)}\label{eq:gamma_hat}
\end{align}
where Step (c) is obtained by dividing \eqref{eq:psigamma_hat2} with \eqref{eq:psigamma_hat1},  and Step (d) is obtained by sum of squares of \eqref{eq:psigamma_hat1} and \eqref{eq:psigamma_hat2}. Both estimates $\hat{\psi}$ and $\hat{\gamma}$ are obtained by averaging their individual estimates over $k\in\mathbb{K}\setminus\{0\}$.
As stated earlier, the estimation of orientation parameters requires the estimates of location parameters. In particular, 
$\hat{\psi}$ given in \eqref{eq:psi_hat} depends on  ($\hat{\theta},\hat{\phi}$), whereas $\hat{\gamma}$ given in \eqref{eq:gamma_hat} depends on  ($\hat{r},\hat{\theta},\hat{\phi}$).  

The proposed 5D pose estimation approach is summarized in Algorithm \ref{alg:5d_pose_estimation}.

\section{Numerical Results and Discussion}\label{sec:results}\vspace{-0.5mm}
This section presents the  normalized mean-square error (NMSE)  performance of Algorithm \ref{alg:5d_pose_estimation} for system parameters as the number of BS antenna $M=9$, number of UE antennas $K=11$, number of RIS elements along $x$ and $y$ axes $N_x = N_y = \sqrt{N}$, number of RIS configurations $P=N$, number of transmissions $L=50$, operational wavelength $\lambda = 0.33$ m, antenna spacing $d_u = d_b = \lambda/2$ m and $d_x = d_y = \lambda/4$ m,
transmission power $P_T=40$ dBm; unless mentioned otherwise. The NMSE performance is evaluated using the Monte Carlo simulations for 10,000 5D-pose realizations of the UE with $\theta \in [10^{\circ},170^{\circ}]$, $\phi \in [10^{\circ},80^{\circ}]$, $\psi \in [15^{\circ},170^{\circ}]$, $\gamma \in [15^{\circ},80^{\circ}]$, and \mbox{\small{$r\in \left[0.62 \lambda^{-1/2} \left(a_R^2+b_R^2\right)^{3 / 4},2\lambda^{-1}\left(a_R^2+b_R^2\right)\right]$}} \cite{Yijin_2023} such that $a_R = 2\nxt dx$ and $b_R = 2\nyt dy$ are the dimensions of the RIS along $x$ and $y$ axes.  
These ranges of $(\theta,\phi,\psi,\gamma)$ are selected to  exclude the boundary issues.

\begin{figure}
    \vspace{-.4cm}
    \hspace{-.3cm}
    \subfloat[Location parameters\label{2a}]{\includegraphics[width=0.25\textwidth]{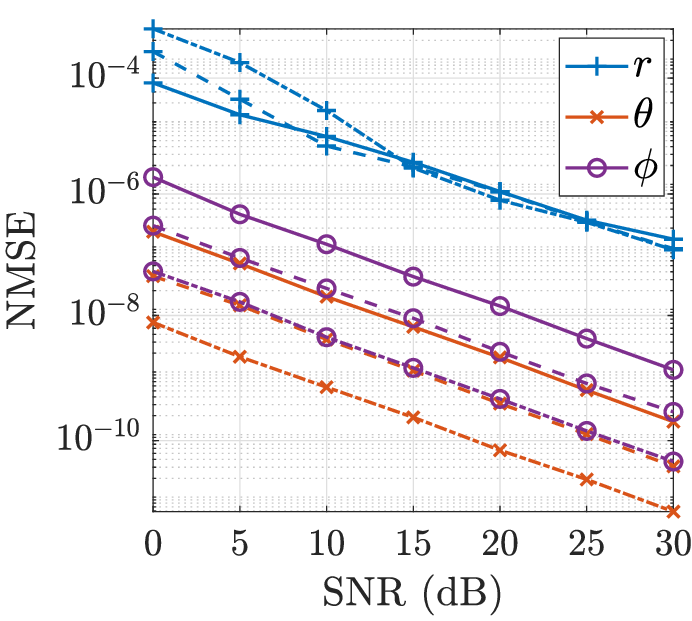}}
    \subfloat[Orientation parameters\label{2b}]{\includegraphics[width=0.25\textwidth]{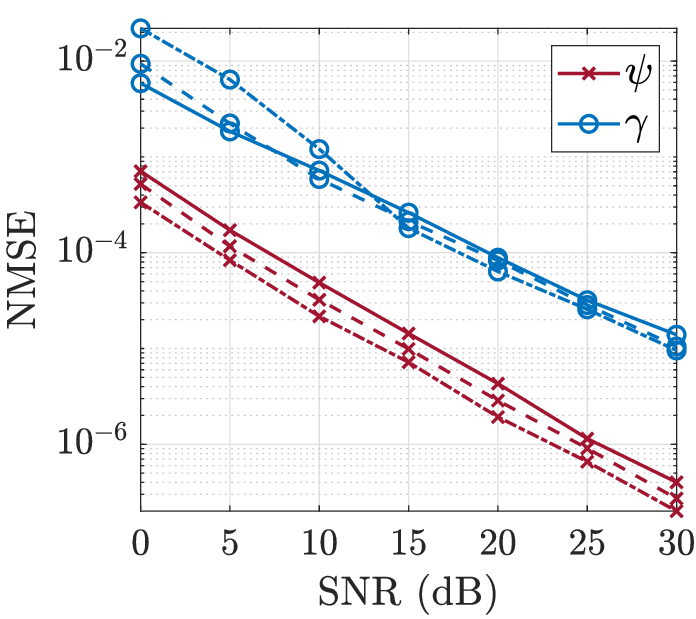}}
    \caption{NMSE vs SNR. The solid, dash and dot-dash curves correspond to $N=121$, $225$ and $441$, respectively.}\vspace{-2.5mm}
    \label{fig:fig_2}
\end{figure}

\begin{figure}
    \vspace{-.3cm}
    \hspace{-.3cm}
    \subfloat[Location parameters\label{3a}]{\includegraphics[width=0.25\textwidth]{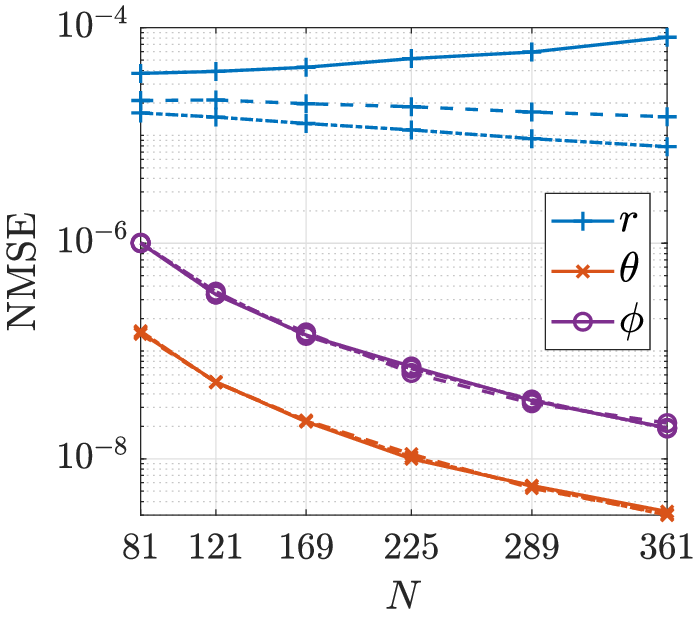}}
    \subfloat[Orientation parameters\label{3b}]{\includegraphics[width=0.25\textwidth]{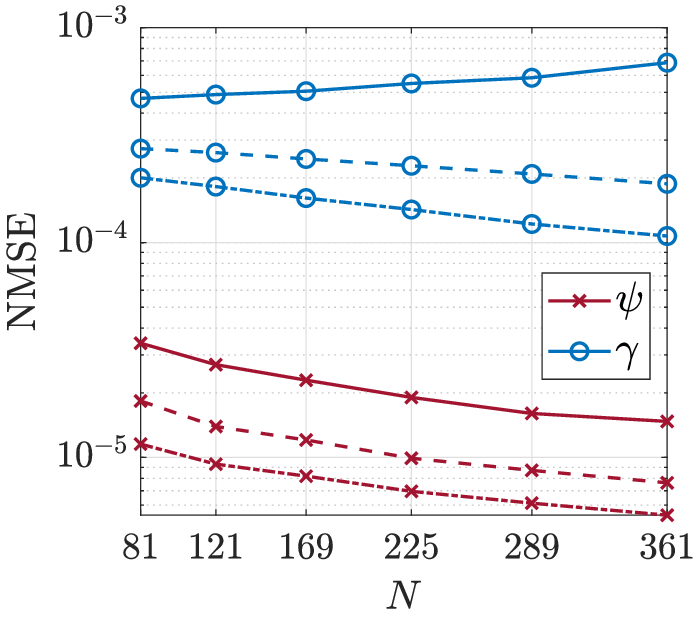}}
    \caption{NMSE of the user pose for $N \in [81,361]$. The plane, dash and dot-dash curves correspond to $K=7$, $11$ and $15$, respectively.}\vspace{-2mm}
    \label{fig:fig_3}
\end{figure}

Fig. \ref{fig:fig_2} shows that the NMSE performance of both location ($r,\theta,\phi$)  and orientation ($\psi,\gamma$) parameters improves with the increase in SNR, as expected.  Fig. \ref{2a} demonstrates that the performance of  $\theta$ and $\phi$ parameters improves with the increase of number of RIS elements $N$. 
On the contrary, the distance parameter $r$ performs poorly with the increase in $N$ at low SNR.
This is because  the near-field region, i.e. range of $r$, grows with the increase in RIS size. 
However,  it is interesting to observe that increasing $N$ allows the estimation of larger span of $r$ without  compromising its NMSE performance at high SNR. 
Fig. \ref{2b} shows that the NMSE performance of orientation-azimuth angle $\psi$ is significantly better compared to that of orientation-elevation angle $\gamma$.  
This is because the estimation of $\gamma$ requires the estimates of ($r,\theta,\phi$), whereas the estimation of $\psi$ requires the estimates of ($\theta,\phi$). Therefore,  the performance of $\gamma$ follows a similar trend to that of  its  limiting factor $r$ and the performance of $\psi$ follows the trend similar to that of   $(\theta,\phi)$. 
Further, for the similar reason, we can observe that performances of   $\psi$ and $\gamma$ are poor as compared to  $\theta$ and $\phi$. 
Furthermore, the performances of $\psi$ and $\gamma$ are not much affected by $N$, as can  also be verified from Fig. \ref{3b}.

\begin{figure}[t]
    \vspace{-.3cm}
    \hspace{-.3cm}
    \subfloat[Location parameters\label{4a}]{\includegraphics[width=0.25\textwidth]{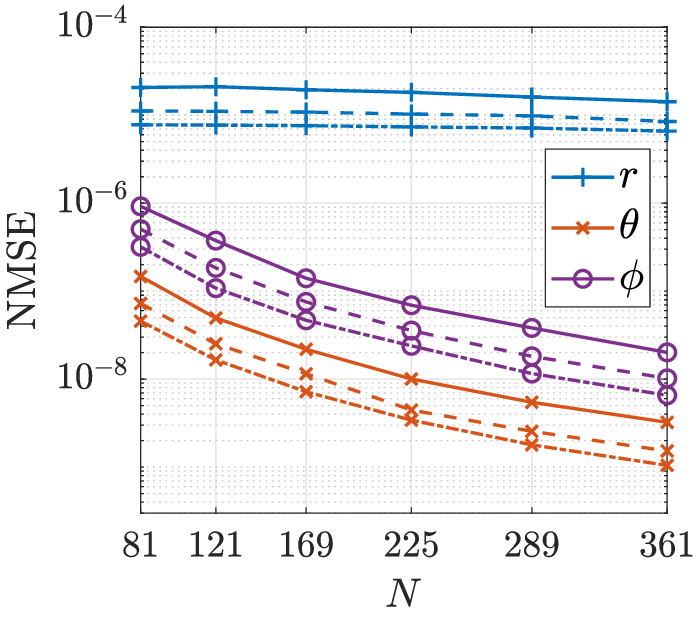}}
    \subfloat[Orientation parameters\label{4b}]{\includegraphics[width=0.25\textwidth]{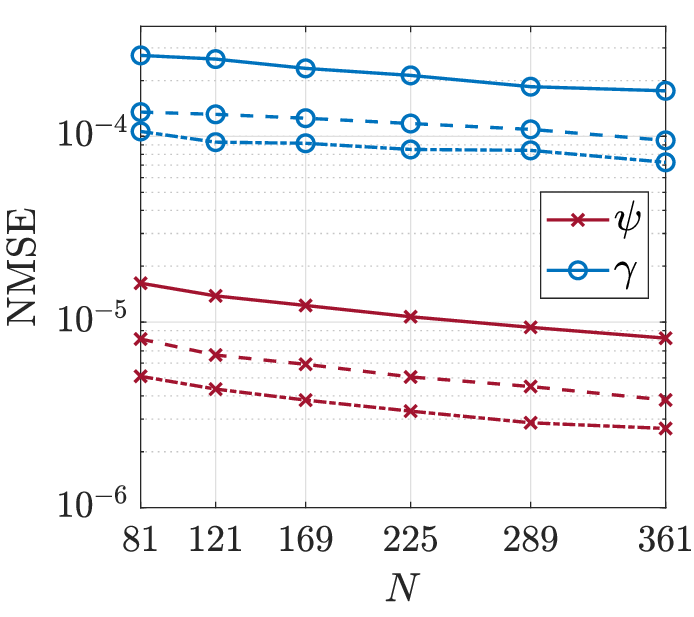}}
    \caption{NMSE of the user pose for $N \in [81,361]$. The plane, dash and dot-dash curves correspond to $P=N$, $2N$ and $3N$, respectively.}\vspace{-2mm}
    \label{fig:fig_4}
\end{figure}

Fig. \ref{fig:fig_3} shows the impact of number of RIS elements $N$ on the NMSE performance for the number of UE antennas $K=7,~11$ and $15$ and the SNR equal to $15$ dB. 
Fig. \ref{3a} shows that the  NMSE of $\theta$ and $\phi$ improves with increase in $N$ and is invariant to $K$ which is expected as the location of UE is independent of size of the ULA.
However, with the increase in $N$, the NMSE of $r$ increases for $K=7$ and decreases for $K=11$ and $15$. This degradation at smaller $K$ is due to the fact that the range of $r$ increases with the increase in RIS size. However, it can be compensated by increasing $K$. As stated earlier, Fig. \ref{3b} shows that the orientation  parameters  $\gamma$ and $\psi$ follow performance trends similar to $r$ and ($\theta,\phi$).   Further, it can be seen that increasing $K$ significantly improves  the performances of these orientation parameters.  
Fig. \ref{fig:fig_4} shows that the NMSE of all parameters improves with the  increase of the number of RIS configurations $P$. This is expected since increasing $P$ means more number of observations and it also means   receiving signal under larger number of  RIS configurations, which essentially increases the chances of capturing  signal with high SNR. Besides, Fig. \ref{fig:fig_4} also shows that the NMSE gain diminishes with increasing $P$.
\vspace{-0mm}
\section{Conclusion}\label{sec:conclusion}  
This paper proposed a low-complexity near-field 5D pose estimation algorithm for RIS-assisted uplink MIMO systems. The proposed approach decouples this 5D problem into five 1D sub-problems by employing suitable channel matrix transformations that leverage the  geometric arrangement of antenna arrays of RIS and UE. Next, these transformed matrices are utilized to model the estimation sub-problems in TLS-ESPRIT form, 
which in turn provide  closed-form solutions to all the parameters. 
Through extensive numerical results, we investigated the accuracy of the proposed approach  and demonstrated that it achieves significantly low NMSE for wide range of system design  parameters.

\vspace{-0mm}
\bibliographystyle{IEEEtran}  
\input{main.bbl}

\end{document}